\documentclass[iop,numberedappendix,twocolappendix]{emulateapj}

\usepackage{natbib}
\usepackage{amsmath}
\usepackage{graphicx}
\usepackage{dcolumn}
\usepackage{bm}
\renewcommand{\phi}{\varphi}
\renewcommand{\theta}{\vartheta}

\shorttitle{The Diffusion and Telegraph Approximation for Solar Energetic Particle Transport}
\shortauthors{Effenberger \& Litvinenko}

\begin{document}

\title{The Diffusion Approximation vs. the Telegraph Equation for
  Modeling Solar Energetic Particle Transport with Adiabatic
  Focusing. I. Isotropic Pitch-angle Scattering}

\author{Frederic Effenberger\altaffilmark{1} and Yuri
  E. Litvinenko\altaffilmark{1}}

\altaffiltext{1}{Department of Mathematics, University of Waikato,
  P.B. 3105, Hamilton, New Zealand}

\begin{abstract}
The diffusion approximation to the Fokker-Planck equation is commonly
used to model the transport of solar energetic particles in
interplanetary space. In this study, we present exact analytical
predictions of a higher order telegraph approximation for particle
transport and compare them with the corresponding predictions of the
diffusion approximation and numerical solutions of the full
Fokker-Planck equation. We specifically investigate the role of the
adiabatic focusing effect of a spatially varying magnetic field on an
evolving particle distribution.  Comparison of the analytical and
numerical results shows that the telegraph approximation reproduces
the particle intensity profiles much more accurately than does the
diffusion approximation, especially when the focusing is strong. 
However, the telegraph approximation appears to offer no significant
advantage over the diffusion approximation for calculating the
particle anisotropy. The telegraph approximation can be a useful tool
for describing both diffusive and wavelike aspects of the cosmic-ray
transport.
\end{abstract}



\section{Introduction}
Cosmic-ray transport remains a subject of intense research activity.
Space weather forecasting relies heavily on models for the solar
energetic particle (SEP) transport in interplanetary space and the
resulting intensities at Earth \citep[e.g.][and references
  therein]{Shea-Smart-2012}. Analysis of the measured SEP profiles
can also yield information on the properties of the medium through
which the particles travel.

The Fokker-Planck equation is typically used in the description of the
evolution of the particle distribution function \citep[see, e.g.][for
  a recent derivation]{Schlickeiser-2011}.  The Fokker-Planck
description of the SEP transport incorporates various important
effects, such as turbulent pitch-angle scattering and adiabatic
focusing due to large-scale gradients in a background magnetic field,
for instance in the Parker spiral field.

To solve the Fokker-Planck equation, analytical approximations or
numerical methods are usually required. In particular, the diffusion
approximation leads to an advection-diffusion equation for the
isotropic part of the distribution.  The equation is known to
approximate the Fokker-Planck equation when the pitch-angle scattering
is strong enough to ensure that the scale of density variation is much
greater than the particle mean free path
\citep{Jokipii-1966,Earl-1974,Earl-1981,Beeck-Wibberenz-1986,Schlickeiser-Shalchi-2008}. 

A shortcoming of the diffusion approximation is an infinite signal
propagation speed that leads to causality violation.  An improved
description of the SEP transport is provided by the telegraph equation
that is consistent with causality. 
\citet{Fisk-Axford-1969} derived the telegraph equation and analyzed 
SEP anisotropies in a bi-directional scattering model. 
Later a modified telegraph equation has
been derived from the Fokker-Planck equation by perturbation methods
\citep{Earl-1976,Gombosi-etal-1993,Schwadron-Gombosi-1994,Pauls-Burger-1994}. 

\citet{Earl-1976} presented a modified telegraph equation for 
the focused particle transport in a spatially varying
magnetic field. The equation, however, described 
the coefficient of an eigenfunction expansion rather than the particle 
density that is the physical quantity of interest. Recently, 
\citet{Litvinenko-Noble-2013} applied a new technique to derive the
telegraph equation for the particle density in a spatially varying
magnetic field of an arbitrary constant focusing strength.  The
technique could be used only for the isotropic pitch-angle scattering,
but \citet{Litvinenko-Schlickeiser-2013} gave a complementary
derivation for an arbitrary scattering rate in a weak focusing limit.

Analytical solutions of the diffusion approximation are employed in
the analysis of spacecraft data \citep{Artmann-etal-2011}.  The
telegraph equation is also amenable to analytical treatment, so it is
natural to ask whether the telegraph equation furnishes a more
accurate description of the SEP transport than the diffusion
approximation.  To address this question, in this paper we follow
\citet{Litvinenko-Noble-2013} and consider a simple but still
physically sensible model of isotropic pitch-angle scattering and
adiabatic focusing with a constant focusing length of a guiding
magnetic field. This enables us to assess the accuracy of the telegraph
approximation using an analytical solution to the modified telegraph equation. 

Our goal is to compare analytical solutions to the diffusion
and telegraph equations and numerical solutions to the full
Fokker-Planck equation, obtained by means of stochastic simulations.
We extend the analytical results in
\citet{Litvinenko-Schlickeiser-2013} by calculating the solution of an
initial value problem of SEP transport, and we extend the numerical
results of \citet{Litvinenko-Noble-2013} by computing both space- and
time-profiles of particle intensities for different parameters, as
well as the anisotropy of the particle distribution.

In the remainder of the paper, we first summarize the results of the
diffusion approximation and the corresponding expressions for the
telegraph approximation. Subsequently, we briefly describe the
numerical scheme, which is similar to the one described in
\citet{Litvinenko-Noble-2013}. Finally, we present and discuss our
results.


\section{Analytical considerations}
\subsection{Basic equations}
The Fokker-Planck equation (which is also
often referred to as focused transport equation) for the distribution
function $f_0=f_0(z,\mu,t)$ of energetic particles is given by \citep[e.g.][]{Roelof-1969,Earl-1981}
\begin{equation}
  \frac{\partial f_0}{\partial t}
+ \mu v \frac{\partial f_0}{\partial z}
+ \frac{v}{2L}(1-\mu^2)\frac{\partial f_0}{\partial \mu}
= \frac{\partial}{\partial \mu} \left( D_{\mu\mu}\frac{\partial f_0}{\partial \mu}\right) . 
\label{eq:FPE}
\end{equation}
Here $f_0$ is the distribution function of energetic particles
(gyrotropic phase-space density), $t$ is time, $\mu$ is the
cosine of the particle pitch angle, $v$ is the (constant) particle
speed, $z$ is the distance along the mean magnetic field $B$,
$L=-B/(\partial B/\partial z)$ is the adiabatic focusing length,
and $D_{\mu\mu}$ is the Fokker-Planck coefficient for pitch-angle
scattering. We consider isotropic pitch-angle scattering:
\begin{equation}
D_{\mu\mu}=D_0(1-\mu^2),
\label{eq:Dmumu}
\end{equation}
where $D_0=$ const.  \citet{Shalchi-etal-2009} analyzed the physical
regimes that lead to isotropic pitch-angle scattering.  We also assume
a constant focusing length $L$ (see, however, the discussion in the
Appendix), and we neglect momentum diffusion, advection with the solar
wind, and particle drift effects.

A mathematically equivalent 
description can be given in terms of the linear density $f (z,\mu,t)$ 
\citep{Earl-1981}, defined by
\begin{equation}
f = \exp(z/L)f_0 . 
\label{eq:lindens}
\end{equation}
The resulting implicit form of the Fokker-Planck equation is 
used below to obtain a stochastic numerical solution. To simplify the comparison 
of the analytical and numerical results, in what follows we express the analytical 
solutions of the diffusion and telegraph equations in terms of 
an isotropic linear density as well.

\subsection{The diffusion approximation}
We begin by summarizing some results for the
diffusion approximation. In this approximation, the equation for the
isotropic particle density
\begin{equation}
F_0 (z,t) = \frac{1}{2}\int_{-1}^{1}f_0d\mu 
\end{equation}
reduces to an advection-diffusion equation 
\citep[see, e.g.,][]{Beeck-Wibberenz-1986}:
\begin{equation}
  \frac{\partial F_0}{\partial t} - u  \frac{\partial F_0}{\partial z} = 
  \kappa_{\parallel} \frac{\partial^2F_0}{\partial z^2} ,
\label{eq:diffusion-advection}
\end{equation}
where $u=\kappa_{\parallel}/L$ is the coherent speed and
$\kappa_{\parallel}$ is the parallel diffusion coefficient. 

The isotropic linear density, defined as the number
of particles per line of force per unit distance parallel to $B$, 
is given by 
\begin{equation}
F (z,t) = \frac{1}{2}\int_{-1}^{1}e^{z/L}f_0d\mu = \exp(z/L) F_0 . 
\label{eq:isotropiclindens}
\end{equation}
Note that the particle conservation is conveniently expressed as 
$N(t) = 2 \int F dz = \mbox{const}$.
Now the fundamental solution to Eq. (\ref{eq:diffusion-advection}), 
that is the solution for a delta-functional injection $F_0(z,0) = \delta(z)$, 
yields the linear density profile 
\begin{equation}
 F(z,t) = \frac{1}{(4\pi \kappa_{\parallel} t)^{1/2}}
 \exp\left[\frac{-(z - ut)^2}{4\kappa_{\parallel}t}\right] .
\label{eq:diffusion-fundamental}
\end{equation}

For isotropic scattering, the parallel diffusion coefficient 
is given by \citep{Beeck-Wibberenz-1986}
\begin{equation}
\kappa_{\parallel} = \lambda_0v\left(\frac{\coth\xi}{\xi}-\frac{1}{\xi^2}\right) ,
\label{eq:kappapar}
\end{equation}
where we have introduced the focusing parameter $\xi=\lambda_0/L$
and the scattering mean free path in the absence of focusing
\citep{Hasselmann-Wibberenz-1970}:
\begin{equation}
\lambda_0 = \frac{3v}{8}\int_{-1}^{1}\frac{(1-\mu^2)^2}{D_{\mu\mu}}d\mu = \frac{v}{2D_0} .
\label{eq:meanfreep}
\end{equation}
A well-known expression for the parallel diffusion coefficient is recovered 
in the limit of no focusing ($\xi \rightarrow 0$):
\begin{equation}
\kappa_{\parallel,0} =\frac{1}{3}\lambda_0v=\frac{v^2}{6D_0} .
\label{eq:nofocus}
\end{equation}

\subsection{The telegraph approximation}
The (modified) telegraph equation for SEP transport is given by 
\begin{equation}
  \frac{\partial F_0}{\partial t}
+ \tau\frac{\partial^2 F_0}{\partial t^2}
= \kappa_\parallel\frac{\partial^2 F_0}{\partial z^2}
+ \xi\kappa_\parallel \frac{\partial F_0}{\partial z} 
\label{eq:telegraph}
\end{equation}
\citep[see, e.g.,][and references therein]{Litvinenko-Noble-2013}. 
Here and in what follows, we use dimensionless variables 
by measuring distances in units of the mean free path $\lambda_0=v/2D_0$, 
speed in units of the constant particle
speed $v$, and time in units of $\lambda_0/v = 1/2D_0$. 

Although we formally recover the diffusion approximation by setting $\tau=0$, 
in practice $\tau$ is not negligibly small. 
As shown in \citet{Litvinenko-Noble-2013}, 
for isotropic scattering the telegraph equation
(\ref{eq:telegraph}) is valid for an arbitrary focusing strength $\xi$, and
$\kappa_\parallel$ and $\tau$ are given by 
\begin{align}
\kappa_\parallel &= \frac{\coth\xi}{\xi} - \frac{1}{\xi^2}, \\
\tau            &= \frac{\tanh\xi}{\xi} . 
\label{eq:kappaandtau}
\end{align}
Consequently $\kappa_\parallel \approx 1/3$ and $\tau \approx 1$ 
in the weak focusing limit $\xi^2 \ll 1$. 

Now consider the initial value problem
\begin{equation}
F_0(z,0) = \delta(z), \;\; \partial_t F_0(z,0)=0 . 
\label{eq:initialprob}
\end{equation}
Here, as in the previous section, 
the distribution function is normalized to unity for simplicity. 
The solution is given by
\begin{equation}
F_0(z,t) = \tau\partial_tG_0 + G_0 , 
\label{eq:initiasol}
\end{equation}
where $G_0$ is a slight generalization of the fundamental solution
given by Eqs. (26) and (27) in
\citet{Litvinenko-Schlickeiser-2013}:
\begin{equation}
G_0(z,t)=\frac{1}{2\sqrt{\kappa_\parallel\tau}}\exp\left(-\frac{\xi z}{2}-\frac{t}{2\tau}\right)I_0(s)
\label{eq:G0}
\end{equation}
for $|z| < t\sqrt{\kappa_\parallel/\tau}$, and zero otherwise. 
$I_0$ is a modified Bessel function of the first kind, and its argument is
\begin{equation}
s = \frac{1}{2}\sqrt{(1-\xi^2\kappa_\parallel\tau)\left(\frac{t^2}{\tau^2}-\frac{z^2}{\kappa_\parallel\tau}\right)}.
\label{eq:s}
\end{equation}
Note that $(1-\xi^2\kappa_\parallel\tau)/\tau=1$ for isotropic scattering. 
We use the fundamental solution $G_0$ from Eq.~(\ref{eq:G0}) to get
\begin{align}
F_0(z,t) &= \frac{1}{4\sqrt{\kappa_\parallel\tau}}\exp\left(-\frac{\xi z}{2}-\frac{t}{2\tau}\right) \nonumber\\
&\times\left[I_0(s) + (1-\xi^2\kappa_\parallel\tau)\frac{t}{2\tau}\frac{I_1(s)}{s}\right]
\label{eq:F0}
\end{align}
for $|z| < t\sqrt{\kappa_\parallel/\tau}$ and
\begin{align}
F_0(z,t) &= \frac{1}{2}\exp\left(-\frac{\xi z}{2}-\frac{t}{2\tau}\right) \nonumber\\
&\times\left[\delta\left(\sqrt\frac{\kappa_\parallel}{\tau}t-z\right) + \delta\left(\sqrt\frac{\kappa_\parallel}{\tau}t+z\right)\right]
\label{eq:F0b}
\end{align}
otherwise.

As before, the linear density $F$ is related to the isotropic
density $F_0$ by
\begin{equation}
F(z,t) = \exp(\xi z)F_0(z,t) . 
\label{eq:Flindens}
\end{equation}

\begin{figure}
\noindent\includegraphics[width=0.48\textwidth]{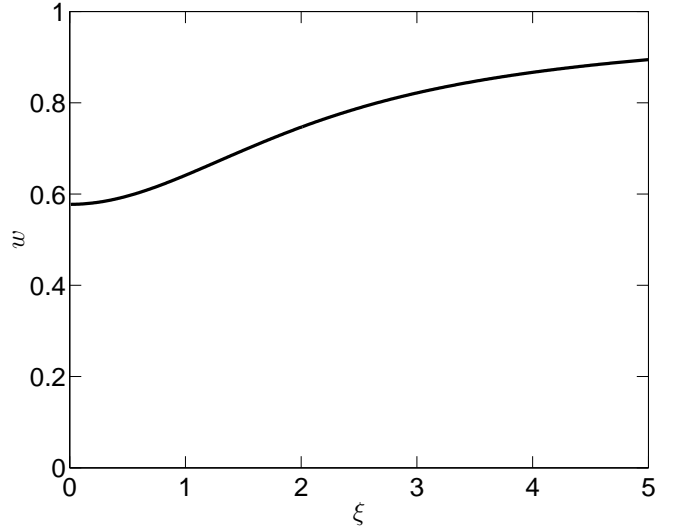}
\caption{Signal propagation speed $w$ of the telegraph equation for different values of $\xi$.}
\label{fig:vsig}
\end{figure}

The dimensionless signal propagation speed
\begin{equation}
w = \sqrt{\frac{\kappa_\parallel}{\tau}}
\end{equation}
of the telegraph equation is plotted in Fig.~\ref{fig:vsig} 
for the case of isotropic scattering. 
In the weak focusing limit $\xi\rightarrow 0$ the propagation speed reduces 
to the value $w=1/\sqrt{3}\approx 0.58$ 
\citep[cf.][]{Earl-1976,Gombosi-etal-1993}.
%


\subsection{Anisotropy}
\label{sec:aniso}
The streaming anisotropy of the particle distribution is defined as
\begin{equation}
A(z,t)=\frac{3\int_{-1}^1\mu f \mathrm{d}\mu}{\int_{-1}^1 f \mathrm{d}\mu} = \frac{3S}{vF_0} ,
\label{eq:streaming-definition}
\end{equation}
where $S$ is the accordingly defined particle flux. \citet{Litvinenko-Schlickeiser-2013} 
calculated the streaming anisotropy in the telegraph approximation (their Eq.~32 
in a slightly different notation):
\begin{equation}
A(z,t)=\frac{1}{F_0}\left(\tau\frac{\partial^2F_0}{\partial t\partial z} - \frac{\partial F_0}{\partial z}\right).
\label{eq:anisotropy}
\end{equation}
In terms of the linear density, the anisotropy is expressed as follows:
\begin{equation}
A(z,t)=\frac{1}{F}\left(\tau\frac{\partial^2F}{\partial t\partial z} - \frac{\partial F}{\partial z}\right) + \xi\left(1-\frac{\tau}{F}\frac{\partial F}{\partial t}\right).
\label{eq:lin-anisotropy}
\end{equation}
In the diffusion approximation, the anisotropy is obtained by 
formally setting $\tau=0$: 
\begin{equation}
A(z,t)=-\frac{1}{F_0}\frac{\partial F_0}{\partial z} , 
\label{eq:diff-anisotropy-formula}
\end{equation}
which, upon inserting the fundamental solution from
Eq.~(\ref{eq:diffusion-fundamental}) gives
\begin{equation}
A(z,t)=\frac{\xi}{2} + \frac{z}{2\kappa_\parallel t} . 
\label{eq:diff-anisotropy}
\end{equation}
%

\section{Stochastic simulation scheme}
Stochastic differential equations are used in many contexts to solve
Fokker-Planck type equations. In space physics, they are often 
employed to solve particle propagation problems, such as cosmic-ray
modulation \citep{Strauss-etal-2011,Effenberger-etal-2012}, SEP
transport \citep{Droege-etal-2010}, shock acceleration
\citep{Achterberg-Schure-2011,Zuo-etal-2011}, 
and pick-up ion evolution \citep{Fichtner-etal-1996,Chalov-Fahr-1998}.  For a
recent account of numerical methods and issues connected to this
approach, see, e.g., \citet{Kopp-etal-2012}. 

The application of the Ito calculus gives a system of stochastic
differential equations, which is completely equivalent to the
Fokker-Planck equation for the linear density, namely \citep{Gardiner-2009}
\begin{align}
dz &= \mu v dt , \\
d\mu &= \left[\frac{v}{2L}(1-\mu^2) - 2D_0\mu\right]dt + \sqrt{2D_0(1-\mu^2)}dW ,
\label{eq:sde}
\end{align}
where $W(t)$ represents a Wiener process with zero mean and variance $t$.

We nondimensionalize this system of equations and solve it numerically, 
using the Milstein approximation scheme
\citep{Litvinenko-Noble-2013,Kloeden-Platen-1995}:
\begin{align}
z_{t+\Delta t} &= z_t + \mu_t\Delta t , 
\label{eq:sdemil1} \\
\mu_{t+\Delta t} &=  \left[\frac{1}{2}\xi(1-\mu_t^2) - \mu_t\right]\Delta t + \sqrt{\Delta t(1-\mu_t^2)}\epsilon_t \nonumber\\ &- \frac{1}{2}\mu_t\Delta t (\epsilon_t^2-1) ,
\label{eq:sdemil2}
\end{align}
where $\epsilon_t$ is a normal random variable with zero mean and
unity variance. We use reflecting boundaries at $\mu=\pm 1$ to
conserve the probability. The following comparisons with the
approximate analytical solutions are performed by simulating a large
number of pseudo-particle orbits according to the above scheme and
obtaining the distribution functions by corresponding averages over
the particle positions. 

We used an isotropic initial pitch-angle distribution in our
simulations. Although the SEP injection can be non-isotropic, the
influence of the initial condition is insignificant after a brief
transitional period of a few scattering times \citep[see a recent
discussion in][and in particular their Figures 3 and
4]{Litvinenko-Noble-2013}.  We verified independently that the results
presented in the following section are only slightly altered if the
initial pitch-angle distribution is proportional to a delta function.
%
 
\section{Results}
\subsection{Spatial intensity behavior}
\begin{figure}
\noindent\includegraphics[width=0.48\textwidth]{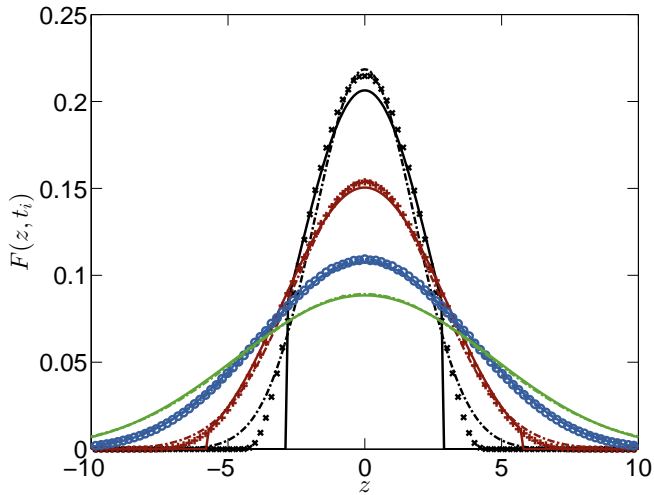}
\caption{Isotropic linear density $F (z, t_i)$ at four different
  times, namely $t_1=5$ (black, 'x'), $t_2=10$ (red, '+'), $t_3=20$ (blue, 'o') and
  $t_4=30$ (green, '.') in the case of no focusing ($\xi=0$). The solid
  lines show the solution of the telegraph equation, given by
  Eq.~(\ref{eq:F0}). The dot-dashed lines give the solution of the
  diffusion approximation (Eq.~\ref{eq:diffusion-fundamental}). The
  symbols show the numerical results, obtained by iterating
  Eqs.~(\ref{eq:sdemil1}) and (\ref{eq:sdemil2}), 
  i.e. the full focused transport problem, with
  $10^7$ particles starting at the origin in each run, and averaging
  without regard to the pitch-angle of the particles.}
\label{fig:F-spaceprofile-xi0}
\end{figure}
To assess the range of validity of the telegraph equation, we
performed stochastic simulations of the type described in the
preceding section with $10^7$ particles starting at the origin in each
run. We then binned the particles in intervals of length 0.1 and
normalized with respect to the number of particles to get a spatial
profile of the particle distribution function (linear density). 
We compared the results with the analytical solution of the telegraph
equation, given by Eqs. (\ref{eq:Flindens}) and (\ref{eq:F0}),
evaluated for different times $t_i$.
Fig.~\ref{fig:F-spaceprofile-xi0} shows the results at four different
times in the case of no focusing ($\xi=0$). A good agreement between
the telegraph solution and the stochastic simulation is found,
especially at later times. For comparison we also show the solution in
the diffusion approximation (Eq.~\ref{eq:diffusion-fundamental}),
which is equally good in this case \citep[see
  also][]{Kota-etal-1982}. Note, however, a slight overshoot of the
diffusion solution at the early time ($t_1=5$), indicating the
non-causality. At $t = t_1 = 5$, no particle 
could have traveled farther from the origin than $z = v t_1 = 5$. 
The telegraph solution, on the other hand, somewhat
underestimates the intensity at larger distances at early times, due
to its lower signal propagation speed $w<v$ (Fig.~\ref{fig:vsig}).

\begin{figure}
\noindent\includegraphics[width=0.48\textwidth]{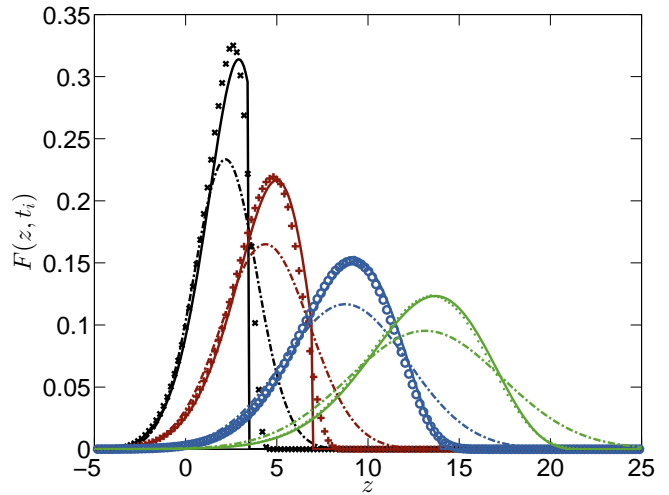}
\caption{Similar to Fig.~\ref{fig:F-spaceprofile-xi0} but for the case of strong focusing ($\xi=1.5$).}
\label{fig:F-spaceprofile-xi15}
\end{figure}
Fig.~\ref{fig:F-spaceprofile-xi15} gives similar plots for the
case of strong focusing ($\xi=1.5$). Here, large differences between
the telegraph and the diffusion solution become visible, 
reinforcing the results in \citet{Litvinenko-Noble-2013}. Clearly 
in this case the telegraph approximation reproduces an evolving density pulse  
much better than the diffusion approximation for all times. 
A feature of interest is an asymmetry of the density profile due to 
the finite particle speed: a sharp front, followed by an extended wake.

\subsection{Temporal intensity behavior}
\begin{figure}
\noindent\includegraphics[width=0.48\textwidth]{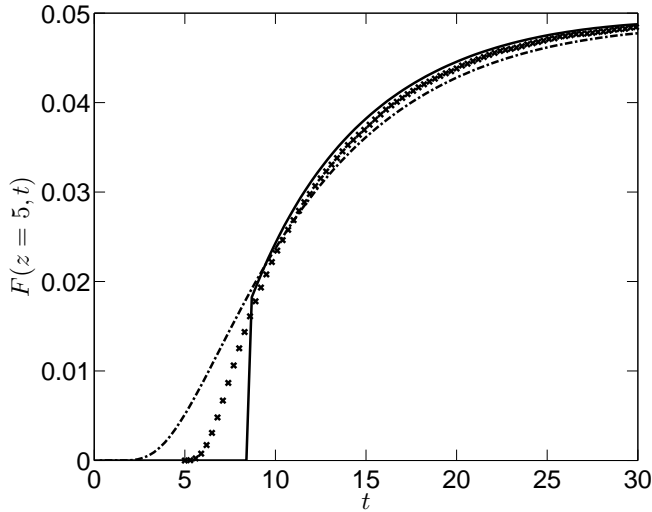}
\caption{Time profile of the isotropic linear density $F$ at a 
  fixed position $z=5$ in the case of no focusing ($\xi=0$). The
  solid line is the analytical solution of the telegraph equation.
  The dot-dashed line gives the solution in the diffusion approximation. 
  The symbols are produced from the same simulations as in
  Fig.~\ref{fig:F-spaceprofile-xi0}.}
\label{fig:F-timeprofile-xi0}
\end{figure}
\begin{figure}
\noindent\includegraphics[width=0.48\textwidth]{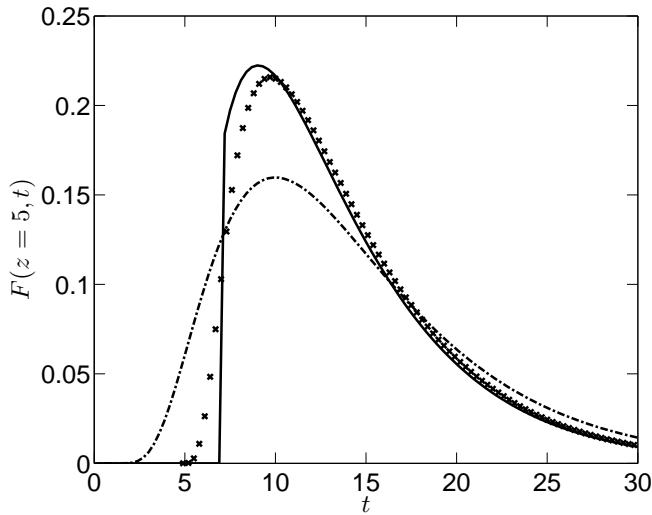}
\caption{Similar to Fig.~\ref{fig:F-timeprofile-xi0} but for the case
  of strong focusing ($\xi=1.5$).}
\label{fig:F-timeprofile-xi15}
\end{figure}
Time profiles of particle intensities are an important tool for 
analyzing the SEP data. Therefore, we extended our comparison to
time profiles at a fixed position. Motivated by the data analysis in 
\citet{Artmann-etal-2011}, we chose $z=5$ and, as previously, 
investigated two cases, namely those of no focusing ($\xi=0$) 
in Fig.~\ref{fig:F-timeprofile-xi0} and strong focusing ($\xi=1.5$) in
Fig.~\ref{fig:F-timeprofile-xi15}. While the diffusion approximation
and the telegraph equation are equally valid in the non-focusing
limit for $t>z/w$, both approximations break down at earlier 
times. By contrast, Fig.~\ref{fig:F-timeprofile-xi15} shows that 
the telegraph equation is much more accurate than the diffusion 
approximation in the strong focusing case. Only at very late times, 
both approximations predict the same value of the intensity.

\subsection{Temporal anisotropy behavior}
\begin{figure}
\noindent\includegraphics[width=0.48\textwidth]{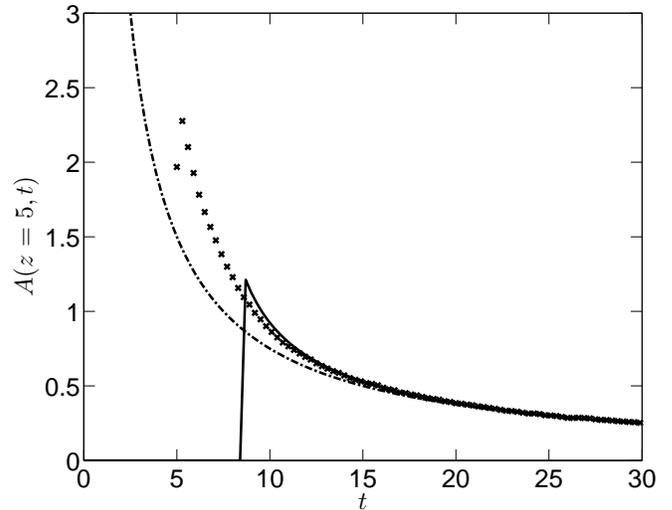}
\caption{Time profile of the anisotropy $A$ at the fixed position
  $z=5$ in the case of no focusing ($\xi=0$). The solid line is
  the numerical solution of Eq.~(\ref{eq:lin-anisotropy}). The
  dot-dashed line gives the solution in the diffusion approximation
  (Eq.~\ref{eq:diff-anisotropy}). The symbols show the numerical
  result for the anisotropy, evaluated from the distribution of
  $10^7$ particles, according to Eq.~(\ref{eq:streaming-definition}).}
\label{fig:A-timeprofile-xi0}
\end{figure}
\begin{figure}
\noindent\includegraphics[width=0.48\textwidth]{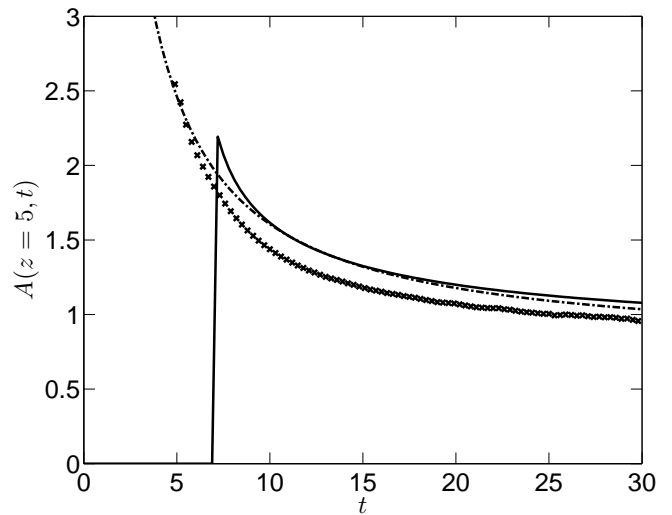}
\caption{Similar to Fig.~\ref{fig:A-timeprofile-xi0} but for the case
  of strong focusing ($\xi=1.5$).}
\label{fig:A-timeprofile-xi15}
\end{figure}
Additional information about energetic particle transport can be obtained 
by analyzing the streaming anisotropy $A(z,t)$ of the observed SEP data. 
We used the stochastic simulation results to compute the anisotropy, 
defined by Eq.~(\ref{eq:streaming-definition}), and we compared 
it with the predictions of the diffusion and telegraph approximations. 
We evaluated equation (\ref{eq:lin-anisotropy}) numerically 
(with a simple finite-difference method), since the analytical 
expressions become quite cumbersome and give no further
insight. 

Figs.~\ref{fig:A-timeprofile-xi0} and \ref{fig:A-timeprofile-xi15}
show the resulting anisotropy profiles at $z=5$ for two cases: no focusing
($\xi=0$) and strong focusing ($\xi=1.5$), respectively. Somewhat
surprisingly, it appears that the accuracy of either approximation is
almost the same, with both predictions slightly overestimating $A$ in
comparison with the numerical results for the case of strong focusing,
even for $t \gg 1$.  The telegraph approximation, however, captures
the early time behavior better for vanishing focusing, although it
cannot accurately model the anisotropy at even earlier times $t <
z/w$. The diffusion approximation can at least give a rough
estimate in the interval $z/v < t < z/w$.  Thus the telegraph
equation, in our parameter range, yields only a slightly better
estimate for $A$ in comparison with the diffusion approximation and
only in a situation of weak or absent focusing.

\section{Discussion}
The telegraph equation approximates a general transport equation
in a number of transport problems, and so it is often desirable to know
how accurate the telegraph approximation is, especially in comparison
with the simpler diffusion approximation
\citep{Gombosi-etal-1993,Porra-etal-1997}.  In this paper we
investigated the validity of the telegraph approximation in a model
problem of solar energetic particle transport in interplanetary space.
We extended recent studies
\citep{Litvinenko-Noble-2013,Litvinenko-Schlickeiser-2013} by
analytically solving an initial value problem for the telegraph
equation, calculating the SEP intensity profiles in space and time,
and comparing the profiles with those obtained from a stochastic
numerical solution of the Fokker-Planck equation.

We conclude that the telegraph approximation reproduces the SEP
intensity profile much more accurately than the diffusion
approximation. The result appears to be related to the finite signal
propagation speed in the telegraph equation, which implies that the
telegraph approximation can describe both diffusive and wavelike
aspects of the intensity evolution. Somewhat surprisingly, we found
that the telegraph approximation offers no significant advantage over
the diffusion approximation for calculating the anisotropy of the SEP
distribution function, with both approximations overestimating the
anisotropy for strong focusing. Consequently, the full Fokker-Planck
equation should be solved in order to determine the pitch-angle
distribution of the energetic particles.

The key simplifying assumptions of the model in this paper are 
the isotropic pitch-angle scattering rate $D_{\mu\mu}$ and a 
constant adiabatic focusing length $L$ of a guiding magnetic field. 
Although $L=\mbox{const}$ is often assumed in theoretical studies 
\citep{Earl-1976,Litvinenko-Schlickeiser-2011}, 
the assumption is valid only as long as the focusing length $L$ 
does not change appreciably over one scattering length $\simeq v/D_0$. 
As we discuss in the Appendix, however, the condition is unlikely to be 
satisfied for the SEP transport close to the Sun. 

In the future we plan to relax both assumptions 
by deriving a more general telegraph-type equation and by 
stochastically simulating the Fokker-Planck equation with 
more realistic $D_{\mu\mu}$ and $L=L(z)$. 
(Note that \citet{Earl-1981} developed a diffusion approximation with $L=L(z)$.)
Further improvements could include more realistic boundary conditions, 
say a reflecting inner boundary. 
Recent studies also emphasized the potential role of 
the observed strong perpendicular transport
\citep{Dresing-etal-2012,Laitinen-etal-2013,Droege-etal-2010}, 
drifts \citep{Marsh-etal-2013}, and modeling of
pitch-angle diffusion with full-orbit methods
\citep[e.g.,][]{Tautz-etal-2013,
  Tautz-2013,Laitinen-etal-2012,Tautz-etal-2012}. 

To sum up, we presented a systematic side-by-side comparison of the
predictions for the SEP transport, made using the diffusion and
telegraph approximations and the Fokker-Planck equation on which the
approximations are based.  We deliberately adopted the simplest
physically meaningful model: isotropic scattering, a constant focusing
length, no advection, momentum diffusion or adiabatic
deceleration. The essential point is that, while various features of
the SEP transport had been previously investigated in detail
numerically
\citep[e.g.,][]{Zank-etal-2000,Lu-etal-2001,Kaghashvili-etal-2004}, we
believe that a simple analytical model for the key features of the
particle transport is valuable since it can guide the numerical
studies.  In the future we intend to relax the simplifying assumptions
of this paper in order to explore the usefulness of the telegraph
approximation more fully.

\begin{acknowledgements} 
  We acknowledge an anonymous referee whose comments motivated us to
  revise parts of the paper.  We thank Horst Fichtner for helpful
  suggestions.
\end{acknowledgements} 
%



\begin{appendix}
\section{The focusing length between the Sun and 1~AU}
\label{sec:parameters}
We discuss the radial dependence of the focusing
length in the Parker interplanetary magnetic field, to quantify the accuracy 
of the assumption of constant focusing for SEPs. The Parker
magnetic field in spherical coordinates ($r,\vartheta,\varphi$) is
given by \citep{Parker-1958}
\begin{align}
B_r &= B_0(r_0)\left(\frac{r_0}{r}\right)^2 , \\
B_{\vartheta} &= 0 , \\
B_{\varphi}   &= -B_r \frac{r\Omega}{u_{sw}} \sin\vartheta . 
\label{eq:parker}             
\end{align}
We assume a constant solar wind speed of $u_{sw}=400$~km/s and
a constant angular velocity of the Sun of $\Omega=2\pi/25$d. In the
following, we consider the case $\vartheta=\pi/2$,
i.e. the field in the ecliptic plane.

The total magnetic field strength is given by
\begin{align}
B = \frac{B(r_0)}{r^2}\sqrt{1+(\beta r)^2} ,
\label{eq:parker-mag}             
\end{align}
where we have introduced $\beta=-\Omega/u_{sw}$, which has a value
of $\beta=-1.09$ AU$^{-1}$ for our choice of parameters.

The focusing length $L$ in the Parker spiral field is given by
\begin{align}
\frac{1}{L} &= -\frac{1}{B}\frac{\partial B}{\partial r}\frac{\mathrm{d} r}{\mathrm{d} z} \nonumber\\
 &= \left(-\frac{1}{B}\right)\cdot\left(-B\frac{2 + (\beta r)^2}{r(1 + (\beta r)^2}\right)\cdot\frac{1}{\sqrt{(1 + (\beta r)^2}} , 
\label{eq:L}             
\end{align}
and so 
\begin{align}
L(r) = \frac{r(1 + (\beta r)^2)^{3/2}}{2 + (\beta r)^2} . 
\label{eq:L}             
\end{align}
More details on the derivation of characteristic parameters in the
Parker field can be found, e.g., in \citet{Artmann-2013}.

\begin{figure}
\centering
\noindent\includegraphics[width=0.48\textwidth]{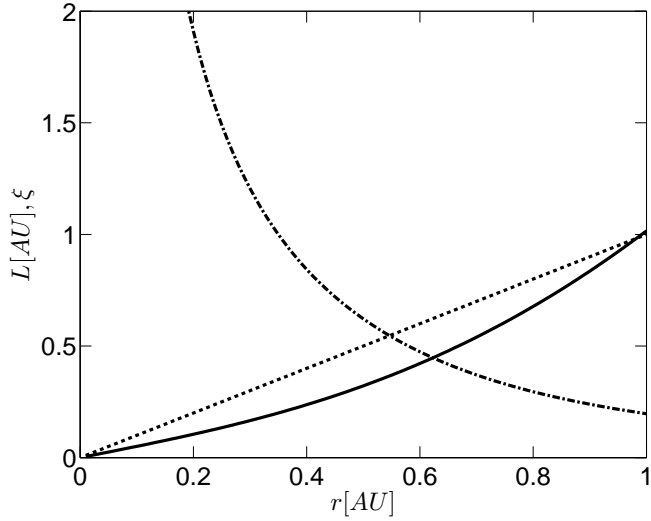}
\caption{Spatial behavior of the focusing length $L$ (solid) and the
  focusing parameter $\xi$ (dot-dashed) in the Parker magnetic field
  between the Sun and 1~AU, assuming a constant mean free path of
  $0.2$ AU. The straight radial line (dotted) is included to guide the
  eye.}
\label{fig:parameters}
\end{figure}

Fig.~(\ref{fig:parameters}) shows the spatial dependence of the focusing
length and the dimensionless focusing parameter $\xi=\lambda_0/L$
between the Sun and 1~AU. We assumed a typical value for the constant
mean free path $\lambda_0$ of 0.2~AU. The focusing length varies 
strongly, and consequently $\xi$ can have both
very large and very small values (depending on the mean free path) between
the Sun and 1~AU.

\end{appendix}


\begin{thebibliography}{43}
\expandafter\ifx\csname natexlab\endcsname\relax\def\natexlab#1{#1}\fi

\bibitem[{{Achterberg} \& {Schure}(2011)}]{Achterberg-Schure-2011}
{Achterberg}, A., \& {Schure}, K.~M. 2011, \mnras, 411, 2628

\bibitem[{{Artmann}(2013)}]{Artmann-2013}
{Artmann}, S. 2013, PhD thesis, Ruhr-University Bochum, Germany

\bibitem[{{Artmann} {et~al.}(2011){Artmann}, {Schlickeiser}, {Agueda},
  {Krucker}, \& {Lin}}]{Artmann-etal-2011}
{Artmann}, S., {Schlickeiser}, R., {Agueda}, N., {Krucker}, S., \& {Lin}, R.~P.
  2011, \aap, 535, A92

\bibitem[{{Beeck} \& {Wibberenz}(1986)}]{Beeck-Wibberenz-1986}
{Beeck}, J., \& {Wibberenz}, G. 1986, \apj, 311, 437

\bibitem[{{Chalov} \& {Fahr}(1998)}]{Chalov-Fahr-1998}
{Chalov}, S.~V., \& {Fahr}, H.~J. 1998, \aap, 335, 746

\bibitem[{{Dresing} {et~al.}(2012){Dresing}, {G{\'o}mez-Herrero}, {Klassen},
  {Heber}, {Kartavykh}, \& {Dr{\"o}ge}}]{Dresing-etal-2012}
{Dresing}, N., {G{\'o}mez-Herrero}, R., {Klassen}, A., {Heber}, B.,
  {Kartavykh}, Y., \& {Dr{\"o}ge}, W. 2012, \solphys, 281, 281

\bibitem[{{Dr{\"o}ge} {et~al.}(2010){Dr{\"o}ge}, {Kartavykh}, {Klecker}, \&
  {Kovaltsov}}]{Droege-etal-2010}
{Dr{\"o}ge}, W., {Kartavykh}, Y.~Y., {Klecker}, B., \& {Kovaltsov}, G.~A. 2010,
  \apj, 709, 912

\bibitem[{{Earl}(1974)}]{Earl-1974}
{Earl}, J.~A. 1974, \apj, 193, 231

\bibitem[{{Earl}(1976)}]{Earl-1976}
---. 1976, \apj, 205, 900

\bibitem[{{Earl}(1981)}]{Earl-1981}
---. 1981, \apj, 251, 739

\bibitem[{{Effenberger} {et~al.}(2012){Effenberger}, {Fichtner}, {Scherer},
  {Barra}, {Kleimann}, \& {Strauss}}]{Effenberger-etal-2012}
{Effenberger}, F., {Fichtner}, H., {Scherer}, K., {Barra}, S., {Kleimann}, J.,
  \& {Strauss}, R.~D. 2012, \apj, 750, 108

\bibitem[{{Fichtner} {et~al.}(1996){Fichtner}, {Le Roux}, {Mall}, \&
  {Rucinski}}]{Fichtner-etal-1996}
{Fichtner}, H., {Le Roux}, J.~A., {Mall}, U., \& {Rucinski}, D. 1996, \aap,
  314, 650

\bibitem[{{Fisk} \& {Axford}(1969)}]{Fisk-Axford-1969}
{Fisk}, L.~A., \& {Axford}, W.~I. 1969, \solphys, 7, 486

\bibitem[{{Gardiner}(2009)}]{Gardiner-2009}
{Gardiner}, C.~W. 2009, {Stochastic Methods: A Handbook for the Natural and
  Social Sciences} (Berlin: Springer)

\bibitem[{{Gombosi} {et~al.}(1993){Gombosi}, {Jokipii}, {Kota}, {Lorencz}, \&
  {Williams}}]{Gombosi-etal-1993}
{Gombosi}, T.~I., {Jokipii}, J.~R., {Kota}, J., {Lorencz}, K., \& {Williams},
  L.~L. 1993, \apj, 403, 377

\bibitem[{{Hasselmann} \& {Wibberenz}(1970)}]{Hasselmann-Wibberenz-1970}
{Hasselmann}, K., \& {Wibberenz}, G. 1970, \apj, 162, 1049

\bibitem[{{Jokipii}(1966)}]{Jokipii-1966}
{Jokipii}, J.~R. 1966, \apj, 146, 480

\bibitem[{{Kaghashvili} {et~al.}(2004){Kaghashvili}, {Zank}, {Lu}, \&
  {Dr{\"o}ge}}]{Kaghashvili-etal-2004}
{Kaghashvili}, E.~K., {Zank}, G.~P., {Lu}, J.~Y., \& {Dr{\"o}ge}, W. 2004,
  Journal of Plasma Physics, 70, 505

\bibitem[{{Kloeden} \& {Platen}(1995)}]{Kloeden-Platen-1995}
{Kloeden}, P., \& {Platen}, E. 1995, {Numerical methods for stochastic
  differential equations} (Spinger, Berlin)

\bibitem[{{Kopp} {et~al.}(2012){Kopp}, {B{\"u}sching}, {Strauss}, \&
  {Potgieter}}]{Kopp-etal-2012}
{Kopp}, A., {B{\"u}sching}, I., {Strauss}, R.~D., \& {Potgieter}, M.~S. 2012,
  Computer Physics Communications, 183, 530

\bibitem[{{Kota} {et~al.}(1982){Kota}, {Merenyi}, {Jokipii}, {Kopriva},
  {Gombosi}, \& {Owens}}]{Kota-etal-1982}
{Kota}, J., {Merenyi}, E., {Jokipii}, J.~R., {Kopriva}, D.~A., {Gombosi},
  T.~I., \& {Owens}, A.~J. 1982, \apj, 254, 398

\bibitem[{{Laitinen} {et~al.}(2012){Laitinen}, {Dalla}, \&
  {Kelly}}]{Laitinen-etal-2012}
{Laitinen}, T., {Dalla}, S., \& {Kelly}, J. 2012, \apj, 749, 103

\bibitem[{{Laitinen} {et~al.}(2013){Laitinen}, {Dalla}, \&
  {Marsh}}]{Laitinen-etal-2013}
{Laitinen}, T., {Dalla}, S., \& {Marsh}, M.~S. 2013, \apjl, 773, L29

\bibitem[{{Litvinenko} \& {Noble}(2013)}]{Litvinenko-Noble-2013}
{Litvinenko}, Y.~E., \& {Noble}, P.~L. 2013, \apj, 765, 31

\bibitem[{{Litvinenko} \& {Schlickeiser}(2011)}]{Litvinenko-Schlickeiser-2011}
{Litvinenko}, Y.~E., \& {Schlickeiser}, R. 2011, \apjl, 732, L31

\bibitem[{{Litvinenko} \& {Schlickeiser}(2013)}]{Litvinenko-Schlickeiser-2013}
---. 2013, \aap, 554, A59

\bibitem[{{Lu} {et~al.}(2001){Lu}, {Zank}, \& {Webb}}]{Lu-etal-2001}
{Lu}, J.~Y., {Zank}, G.~P., \& {Webb}, G.~M. 2001, \apj, 550, 34

\bibitem[{{Marsh} {et~al.}(2013){Marsh}, {Dalla}, {Kelly}, \&
  {Laitinen}}]{Marsh-etal-2013}
{Marsh}, M.~S., {Dalla}, S., {Kelly}, J., \& {Laitinen}, T. 2013, \apj, 774, 4

\bibitem[{{Parker}(1958)}]{Parker-1958}
{Parker}, E.~N. 1958, \apj, 128, 664

\bibitem[{{Pauls} \& {Burger}(1994)}]{Pauls-Burger-1994}
{Pauls}, H.~L., \& {Burger}, R.~A. 1994, \apj, 427, 927

\bibitem[{{Porra} {et~al.}(1997){Porra}, {Masoliver}, \&
  {Weiss}}]{Porra-etal-1997}
{Porra}, J.~M., {Masoliver}, J., \& {Weiss}, G.~H. 1997, \pre, 55, 7771

\bibitem[{{Roelof}(1969)}]{Roelof-1969}
{Roelof}, E.~C. 1969, in Lectures in High-Energy Astrophysics, ed.
  H.~{{\"O}gelman} \& J.~R. {Wayland}, 111

\bibitem[{{Schlickeiser}(2011)}]{Schlickeiser-2011}
{Schlickeiser}, R. 2011, \apj, 732, 96

\bibitem[{{Schlickeiser} \& {Shalchi}(2008)}]{Schlickeiser-Shalchi-2008}
{Schlickeiser}, R., \& {Shalchi}, A. 2008, \apj, 686, 292

\bibitem[{{Schwadron} \& {Gombosi}(1994)}]{Schwadron-Gombosi-1994}
{Schwadron}, N.~A., \& {Gombosi}, T.~I. 1994, \jgr, 99, 19301

\bibitem[{{Shalchi} {et~al.}(2009){Shalchi}, {Koda}, {Tautz}, \&
  {Schlickeiser}}]{Shalchi-etal-2009}
{Shalchi}, A., {Koda}, T.~{\AA}., {Tautz}, R.~C., \& {Schlickeiser}, R. 2009,
  \aap, 507, 589

\bibitem[{{Shea} \& {Smart}(2012)}]{Shea-Smart-2012}
{Shea}, M.~A., \& {Smart}, D.~F. 2012, \ssr, 171, 161

\bibitem[{{Strauss} {et~al.}(2011){Strauss}, {Potgieter}, {B{\"u}sching}, \&
  {Kopp}}]{Strauss-etal-2011}
{Strauss}, R.~D., {Potgieter}, M.~S., {B{\"u}sching}, I., \& {Kopp}, A. 2011,
  \apj, 735, 83

\bibitem[{{Tautz}(2013)}]{Tautz-2013}
{Tautz}, R.~C. 2013, \aap, 558, A148

\bibitem[{{Tautz} {et~al.}(2013){Tautz}, {Dosch}, {Effenberger}, {Fichtner}, \&
  {Kopp}}]{Tautz-etal-2013}
{Tautz}, R.~C., {Dosch}, A., {Effenberger}, F., {Fichtner}, H., \& {Kopp}, A.
  2013, \aap, 558, A147

\bibitem[{{Tautz} {et~al.}(2012){Tautz}, {Dosch}, \&
  {Lerche}}]{Tautz-etal-2012}
{Tautz}, R.~C., {Dosch}, A., \& {Lerche}, I. 2012, \aap, 545, A149

\bibitem[{{Zank} {et~al.}(2000){Zank}, {Lu}, {Rice}, \&
  {Webb}}]{Zank-etal-2000}
{Zank}, G.~P., {Lu}, J.~Y., {Rice}, W.~K.~M., \& {Webb}, G.~M. 2000, Journal of
  Plasma Physics, 64, 507

\bibitem[{{Zuo} {et~al.}(2011){Zuo}, {Zhang}, {Gamayunov}, {Rassoul}, \&
  {Luo}}]{Zuo-etal-2011}
{Zuo}, P., {Zhang}, M., {Gamayunov}, K., {Rassoul}, H., \& {Luo}, X. 2011,
  \apj, 738, 168

\end{thebibliography}

\end{document}